\pdfoutput=1
\documentclass[]{tPHL2e_edit}
\usepackage[final]{pdfpages}

\begin{document}
\doi{10.1080/09500831003800863}
\issn{1362-3036}
\issnp{0950-0839} 
\jvol{90} \jnum{8} \jyear{2010} \jmonth{August}

\markboth{Sven A. E. Johansson and G{\"o}ran Wahnstr{\"o}m}{}


\title{Theory of ultrathin films at metal-ceramic interfaces}

\author{Sven A. E. Johansson$^{\ast}$\thanks{$^\ast$Corresponding author. Email: sven.johansson@chalmers.se
\vspace{6pt}} and G{\"o}ran Wahnstr{\"o}m\\\vspace{6pt}  {\em{Department of Applied Physics, Chalmers University of Technology, SE-412 96 G{\"o}teborg, Sweden}\\\vspace{6pt}\received{\today} }}

\maketitle

\begin{abstract}
A theoretical model for understanding the formation of interfacial thin films is presented, which combines density functional theory calculations for interface energies with thermodynamic modeling techniques for multicomponent bulk systems. The theory is applied to thin film formation in VC-doped WC-Co cemented carbides. It is predicted that ultrathin VC films may exist in WC/Co interfaces at the high temperature sintering conditions where most of the WC grain growth occurs, which provides an explanation of the grain growth inhibiting effect of VC additions in the WC-Co system.\bigskip

\bigskip

\end{abstract}

\section{Introduction}
Ultrathin films are essential for many applications in nanoscale science and technology. Considerable scientific interest is devoted to thin oxide films at metal surfaces \cite{KrScNaShKoVa05,RoReSc07,FrRiSc07}, for which applications are found in catalysis, semiconductor devices as well as mechanical wear and corrosion protection. Correctly assessing the thermodynamic stability of such films is an important issue \cite{Ca06}. New thin film surface alloys have been discovered and the stability of nano-sized patterns in the ultrathin films investigated with respect to strain and chemical effects \cite{ThOzScBaAsHoChHw01,OzAsHo02}. Formation of ultrathin films is also observed in metal-ceramic materials after sintering of nanoparticles into a bulk nanocrystalline material \cite{LaLoDo04}.

In this Letter, we consider an interface between two dissimilar materials, a metal-ceramic interface, and the propensity to form, at the interface, an ultrathin film of a secondary phase, whose corresponding bulk phase is thermodynamically unstable. We develop a model for the thin film formation and stability as function of temperature and show that it can be stabilized by interfacial effects.

For modeling of interfaces with complex bonding such as metal-ceramic interfaces, first-principles calculations in the framework of density functional theory (DFT) has proved to be a very useful tool \cite{Fi96,DuHaLu00,ChWaAlLa05}. However, phase stabilities at high temperatures are cumbersome to predict with sufficient accuracy using DFT \cite{TuAbBuFrGrKaKoMaOhPiScZh07} and therefore we combine DFT calculations with driving forces of nucleation from well-developed thermodynamic modeling techniques \cite{SuJaAn85}. These modeling techniques describe bulk phases but lack information on interfacial effects. 

We apply the developed methodology to WC-Co cemented carbides. These materials are produced by means of powder metallurgy, where powders of carbides and metal Co are sintered together into a hard and dense material with excellent mechanical properties. The materials are used in cutting and wear resistant tools and are of considerable industrial importance. Much effort has been invested in developing nanostructured cemented carbides by reducing the mean WC grain size in the finished product, because a fine and homogeneous microstructure has superior mechanical properties \cite{FaWa08}. To control the microstructure and retain a fine grain size throughout the sintering, dopants in the form of VC, $\text{Cr}_3\text{C}_2$, TiC, TaC or NbC powders are added, of which VC has proved to be the most effective inhibitor in ultrafine alloys \cite{ScBoLu95}. The grain growth inhibiting effect of the dopants is well established, whereas the inhibiting mechanism on the atomic scale is not understood. Experimental high-resolution electron microscopy (HREM) studies \cite{JaYaIkSaTaOkTa98,YaIkSa00,YaIkWaSaTaOkTa01,LaHaThLa02,LaThHaTh03} have shown the existence of V- and Cr-rich films of cubic carbide structure at WC/Co interfaces in VC- and $\text{Cr}_3\text{C}_2$-doped materials, respectively, at low temperatures, after liquid phase sintering and cooling. In order to predict the possible growth inhibiting effect of such films, it is decisive to determine whether these can exist at high temperature liquid phase sintering conditions, where most of the grain growth occurs and where HREM cannot be applied.

\section{Model}

\subsection{General description}

The interface energy $\gamma$ can be expressed as the excess Gibbs energy per unit interface area due to the presence of an interface \cite{SuBa96},
\begin{equation}
	\label{eq:interfaceEnergy}
	\gamma = \frac{1}{A} \left( G - \sum_i n_i \mu_i \right).
\end{equation}
$G$ is the Gibbs energy for the entire considered system, which contains an interface of area $A$. $n_i$ is the number of atoms in the system of component $i$ with a corresponding chemical potential $\mu_i$ determined by the reservoirs with which the system is in equilibrium. In the present case, the reservoirs are stoichiometric WC and a binder phase. The latter consists of mainly Co (and will be denoted ``Co'') but also of dissolved W, C, and M atoms, where M = V, Cr, Ti, etc. depending on the doping.

We consider the propensity to form a thin film consisting of $N$ layers of carbide phase, here denoted MC, at the interface between WC and the binder phase. For sufficiently thick films, the interface can be regarded as split into a WC/MC interface and a MC/Co interface separated by a slab of MC. The energy for such a film-covered interface can then be decomposed as
\begin{equation}
	\label{eq:gamma_film}
	\gamma_\text{film} = \gamma_\text{WC/MC} + \gamma_\text{MC/Co} + N \left(  \Delta \bar{g}_\text{MC} +  \bar{e}_\text{MC} \right),
\end{equation}
where $\gamma_\text{WC/MC}$ ($\gamma_\text{MC/Co}$) is the interface energy between WC and MC (MC and Co). The last term of Eq.~(\ref{eq:gamma_film}) represents the energetic cost per layer of building the MC phase and, thus, increases linearly with the number of layers $N$ of the film. For the corresponding energy per formula unit we use the notation $\Delta g_\text{MC} = S \Delta \bar{g}_\text{MC}$, where $S$ is the interface area of one formula unit. The energy $\Delta g_\text{MC}$ is the negative driving force of nucleation. Assuming a stoichiometric carbide phase, it equals
\begin{equation}
	\Delta g_\text{MC} = g_\text{MC} - \mu_\text{M} - \mu_\text{C},
\end{equation}
where $g_\text{MC}$ is the Gibbs energy per formula unit of MC and $\mu_\text{M}$ ($\mu_\text{C}$) is the chemical potential of M (C). A positive value of $\Delta g_\text{MC}$ thus implies that the MC phase is unstable. We consider coherently strained thin films and the last quantity of Eq.~(\ref{eq:gamma_film}), $\bar{e}_\text{MC}$, is the strain energy per layer needed to bring the film into epitaxy with the underlying WC.

The energy $\gamma_\text{film}$ should be compared with the energy $\gamma_\text{WC/Co}$ of a WC/Co interface not covered by any MC film. The film is stable if $\gamma_\text{film} < \gamma_\text{WC/Co}$ and, hence, a positive value of
\begin{equation}
	\label{eq:deltaGamma}
	\Delta \gamma_\text{MC} = \gamma_\text{WC/Co} - \left( \gamma_\text{WC/MC} + \gamma_\text{MC/Co} \right)
\end{equation}
indicates that film formation is possible. An upper limit of the film thickness $N_\text{limit}$ is found where the energy of the film-covered interface matches the energy of the uncovered interface, $\gamma_\text{film}=\gamma_\text{WC/Co}$, which yields
\begin{equation}
	\label{eq:N_limit}
	N_\text{limit} = \frac{\Delta \gamma_\text{MC}}{\Delta \bar{g}_\text{MC}  + \bar{e}_\text{MC}}.
\end{equation}
If $N_\text{limit}$ is larger than a few atomic layers, the interface may contain a thin film.

\begin{figure}
	\begin{center}
		\includegraphics{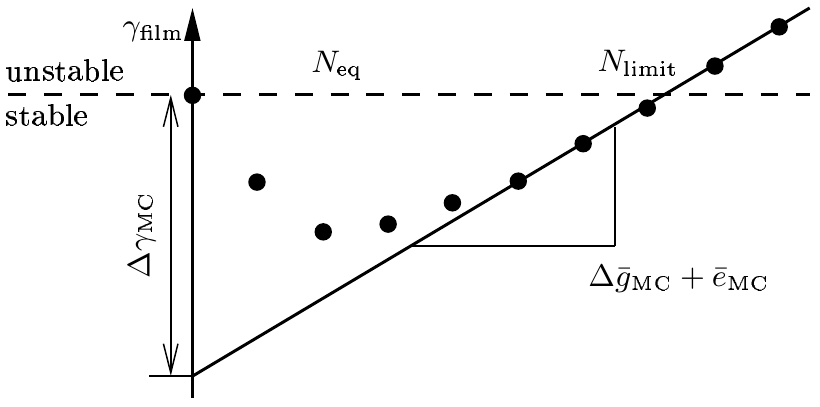}
	\end{center}
	\caption{\label{fig:gamma_film_sch}A schematic illustration of $\gamma_\text{film}$ as function of film thickness $N$. The circles denote the energy in an atomistic treatment and the solid line follows the decomposition in Eq.~(\ref{eq:gamma_film}).}
\end{figure}

The decomposition of $\gamma_\text{film}$ in Eq.~(\ref{eq:gamma_film}) is valid only if the film is sufficiently thick, so that the interior of the film is bulk-like and the interface energies are well-defined. For thin films, $\gamma_\text{film}$ will deviate from the linear dependence on $N$, which implies that $\gamma_\text{film}$ can have a minimum as function of $N$ for a finite film thickness. In equilibrium, this optimal film thickness $N_\text{eq}$ will be attained. A schematic illustration of $\gamma_\text{film}$ is given in Figure~\ref{fig:gamma_film_sch}.

\subsection{Specific model and method}

In the present paper we concentrate on VC-doped WC-Co. Results for other MC dopants will be presented elsewhere \cite{JoWa}. WC has a simple hexagonal lattice structure. After liquid-phase sintering in Co, the WC grains are predominantly delimited by their basal $\left( 0001 \right)$ and prismatic $\left(1 0 \bar{1} 0 \right)$ surfaces \cite{Ex79}. Here, we focus on the basal surface, which can be either W- or C-terminated. We choose the WC surface and the cubic VC phase to be oriented according to
\begin{equation}
	\label{eq:orientation}
	\left( 0001 \right)_\text{WC} \parallel \left( 111 \right)_\text{VC}, \left[ \bar{1} 2 \bar{1} 0 \right]_\text{WC} \parallel \left[  \bar{1} 1 0 \right]_\text{VC}.
\end{equation}
This orientation relationship between WC and MC has been experimentally established for both V- and Cr-doped WC-Co \cite{LaHaThLa02,DeBaPaLaAl04}.

Interface calculations are performed using DFT and the current computational methodology has been developed in a series of papers \cite{ChDuWa02,ChWa03,ChWa04}. Computational results have been compared with experimental data \cite{ChDuWa02,ChWa03} and the use of the generalized gradient approximation (GGA) for the exchange-correlation effects within DFT in interface calculations has been motivated \cite{ChWa04}. Further computational details can be found in \cite{onlinesupp09}.

To model the interfaces, we use standard supercell slab geometries, where the interfaces are made coherent by straining the VC and Co phases to match the WC basal surface. The associated strain energies are minimized by relaxing the interplanar distances to account for the Poisson effect. In the calculation of interface energies, the strain energies are canceled, so that $\gamma_\text{WC/VC}$, $\gamma_\text{VC/Co}$, and $\gamma_\text{WC/Co}$ represent the chemical contribution to the interface energy and do not depend on system size.

To define an absolute value of the interface energy, a definite value of the carbon chemical potential $\mu_\text{C}$ is required. Within the two-phase region (WC+Co) of the W-C-Co system, $\mu_\text{C}$ can be varied in a small interval limited by the formation of either graphite or a metal-carbide phase $\text{(W,Co)}_6\text{C}$ \cite{FrMa08}. We assume carbon to be in equilibrium with graphite and use the corresponding computed value for $\mu_\text{C}$ \cite{ChWaAlLa05}.

\section{Results and discussion}

\subsection{Interface energies}

In the $\left \langle 0001 \right \rangle_\text{WC}$ and $\left \langle 111 \right \rangle_\text{VC}$ directions, the WC and VC phases consist of alternating layers of metal and carbon atoms. Assigning the positions $A$ to $(0,0)$, $B$ to $(\frac{1}{3}, \frac{2}{3})$ and $C$ to $(\frac{2}{3}, \frac{1}{3})$ with respect to hexagonal lattice vectors in the $(0001)_\text{WC}$ plane, the stacking sequence for WC is $\ldots \text{W}_A \text{C}_B \text{W}_A \text{C}_B \ldots$ and for VC $\ldots \text{V}_A \text{C}_C \text{V}_B \text{C}_A \text{V}_C \text{C}_B \ldots$ along $\left [ 0001 \right ]_\text{WC}$ and $\left [ 111 \right ]_\text{VC}$, respectively. We have calculated interface energies for all possible stacking sequences involving the $A$, $B$ and $C$ positions and varying atomic terminations of the phases. Only stoichiometric VC compounds are considered. To assign a value of $\gamma_\text{WC/VC}$, we pick the lowest energy among the different stackings and terminations. The resulting interface energy is found to be slightly negative, $\gamma_\text{WC/VC} = -0.03 \ \text{J/m}^2$ \cite{onlinesupp09}.

The predicted WC/VC interface structure can be compared with experimental findings. By comparing experimental and simulated HREM images of a WC(0001)/VC(111) interface in a VC-doped WC-Co, a particular interfacial stacking has been suggested \cite{LaHaThLo04}. We find that the lowest energy is obtained for $\ldots \text{C}_B \text{W}_A \text{C}_B \text{V}_A \text{C}_C \text{V}_B \text{C}_A \ldots$ \cite{onlinesupp09} which is in complete agreement with the stacking suggested in Ref.~\cite{LaHaThLo04}.

The WC/Co and VC/Co interfaces need to be treated with more concern, since at relevant sintering temperatures, the Co binder is in its liquid phase. A solid-liquid interface energy is usually considered as stemming from two separate parts (see, e.g. \cite{Wa80}): a chemical contribution due to differing atomic bonding at the interface, and a structural contribution due to disturbance of the liquid's structure caused by the interface. One should notice, that only the difference between the two interface energies $\gamma_\text{WC/Co}$ and $\gamma_\text{VC/Co}$ enters into the expression (\ref{eq:deltaGamma}) for the key quantity $\Delta \gamma_\text{VC}$. Therefore, the geometrical similarity of the WC(0001) and VC(111) surfaces implies that structural contributions to $\Delta \gamma_\text{VC}$ will to large extent cancel.

\begin{figure}
	\begin{center}	
		\includegraphics{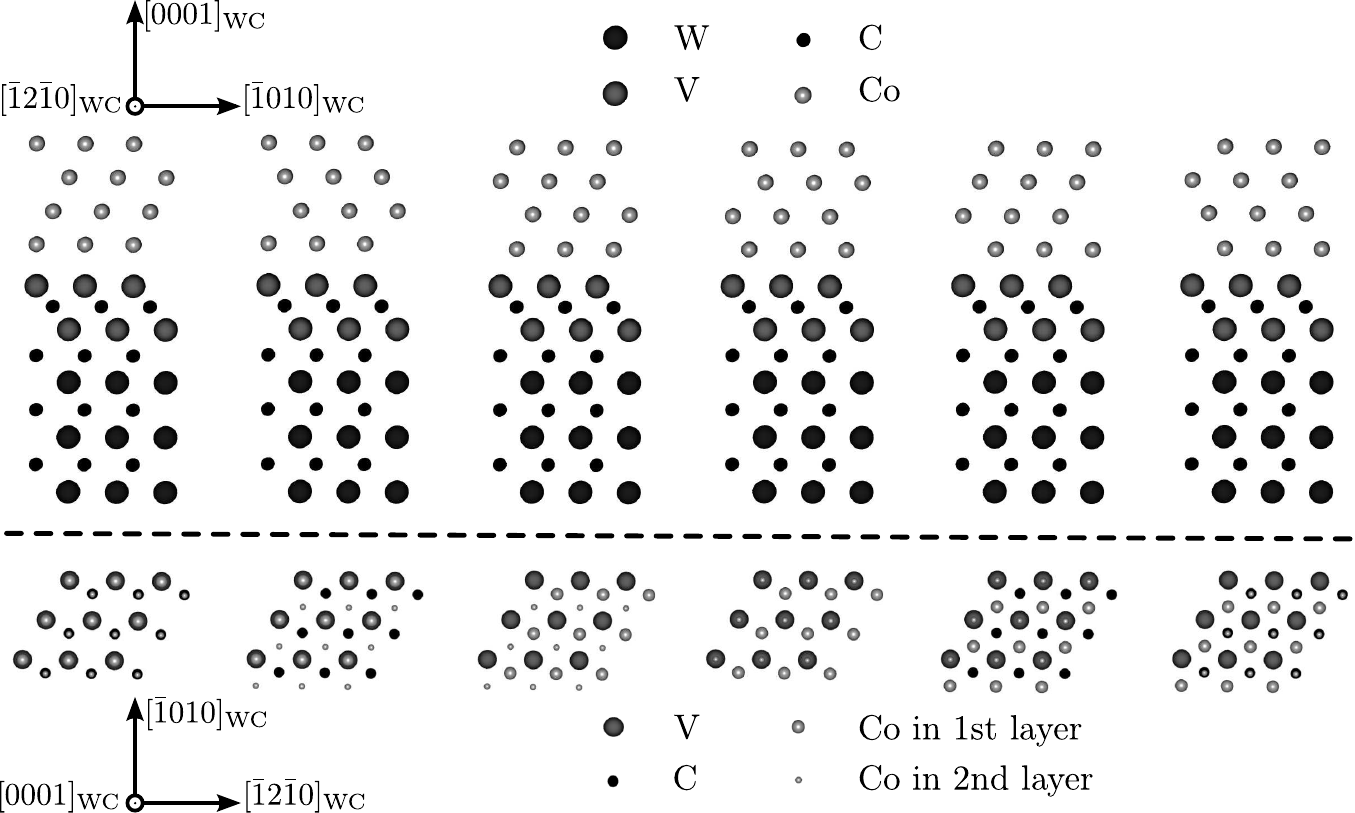}
	\end{center}
	\caption{\label{fig:interface}The optimal atomic structure of a WC/VC/Co interface containing two layers of stoichiometric VC. Top: Atomic positions projected onto a plane with normal $[\bar{1} 2 \bar{1} 0]_\text{WC}$. Bottom: The positions of the V, C and Co atoms closest to the interface projected onto the interface plane. The six depicted atomic configurations of Co are used to calculate a mean value of the carbide/Co interface energy for the thin film configuration. The same set of relative translations and rotations of Co with respect to the topmost carbide layer is also used in calculating WC/Co and VC/Co interface energies.}
\end{figure}

We choose to model the WC/Co and VC/Co interfaces in the same geometry as the WC/VC interface, which means that we are effectively modeling WC(0001)/Co(111) and VC(111)/Co(111) interfaces with different Co stackings. Following the preceding argument, we do not expect the particular choice of Co orientation to influence $\Delta \gamma_\text{VC}$ in any decisive way, although existing experimental observations of this orientation between WC and solid Co \cite{MoSt96_1,BoLaLoMi08} do motivate its use as model geometry. At high temperatures, the motion of liquid Co atoms will constantly change the atomic structure at the interface. Therefore, we use a mean value of energies for different stackings of Co in the interface to approximate $\gamma_\text{WC/Co}$ and $\gamma_\text{VC/Co}$, respectively. In Figure~\ref{fig:interface} we depict the six different atomic arrangements of Co used in calculating the mean energy of the carbide/Co interface. The resulting value for the interface energy can be regarded as a good approximation for an incoherent interface \cite{ChWaLaAl07}. The termination of the WC or VC slabs (metallic or C) is consistently chosen to yield the lowest mean interface energy.

We find that for WC/Co, the interface is C-terminated with an energy $\gamma_\text{WC/Co} = 1.01 \ \text{J/m}^2$  \cite{onlinesupp09}. For VC/Co, the interface is V-terminated with $\gamma_\text{VC/Co} = 0.00 \ \text{J/m}^2$ \cite{onlinesupp09}. The resulting difference in interface energy is $\Delta \gamma_\text{VC} = 1.03 \ \text{J/m}^2$, which we henceforth will assume to be temperature independent.

The final result of $\Delta \gamma_\text{VC}$ does not strongly depend on the choice of averaging of the various Co stackings in the interfaces. For instance, by following the above procedure but disregarding the two most high-energetic configurations in each interface, $\Delta \gamma_\text{VC}$ increases by $0.14 \ \text{J/m}^2$. The impact on $\Delta \gamma_\text{VC}$ due to ordering of the liquid Co atoms in the vicinity of the WC or VC interfaces is therefore expected to be maximally of this size. Temperature-dependent deviations from our estimation of $\Delta \gamma_\text{VC}$ may arise from differences in the vibrational spectrum of V, W, and C atoms at the interface as compared with their respective reference states. Following the procedure of Reuter and Scheffler \cite{ReSc01}, we estimate this effect to be maximally $\pm 0.3 \ \text{J/m}^2$. Considering both these effects, we expect $\Delta \gamma_\text{VC}$ to be determined within an accuracy of $\pm 0.35 \ \text{J/m}^2$. By accounting for a mixed carbide layer instead of the stoichiometric VC compounds considered so far, further entropic effects are introduced (see below).

\subsection{Thin film energy and composition}

In addition to $\Delta \gamma_\text{VC}$, the energetic costs, $\Delta \bar{g}_\text{VC}$ and $\bar{e}_\text{VC}$, to build the VC thin films are needed to predict stability. To compress VC from its calculated equilibrium lattice parameter in the interface, $a_\text{VC}/\sqrt{2} = 2.944 \ \text{\r{A}}$, to the calculated WC lattice parameter, $a_\text{WC} = 2.919 \ \text{\r{A}}$, an energy of only $\bar{e}_\text{VC} = 0.011 \ \text{J/m}^2$ ($0.005 \ \text{eV}$ per VC formula unit) is required.

The driving force of nucleation, $- \Delta g_\text{VC}$, depends sensitively on the temperature and composition of the system. The most stable cubic carbide is not stoichiometric VC, but a mixed $\text{(V,W,Co)C}_x$ phase. In its at $1593 \ \text{K}$ experimentally measured composition, V constitute 75\% and W 23\% of the metallic atoms and the C vacancy concentration is 11\% \cite{FrMa08}. To determine the driving force of nucleation we have used the Thermocalc software \cite{SuJaAn85} with a database for Cr and V additions in the W-C-Co system \cite{FrMa08}. The database has been assessed from experimental observations using the Calphad method and models temperature dependent Gibbs energies of the phases occurring in the W-C-Co system.

\begin{figure}
	\begin{center}
		\includegraphics{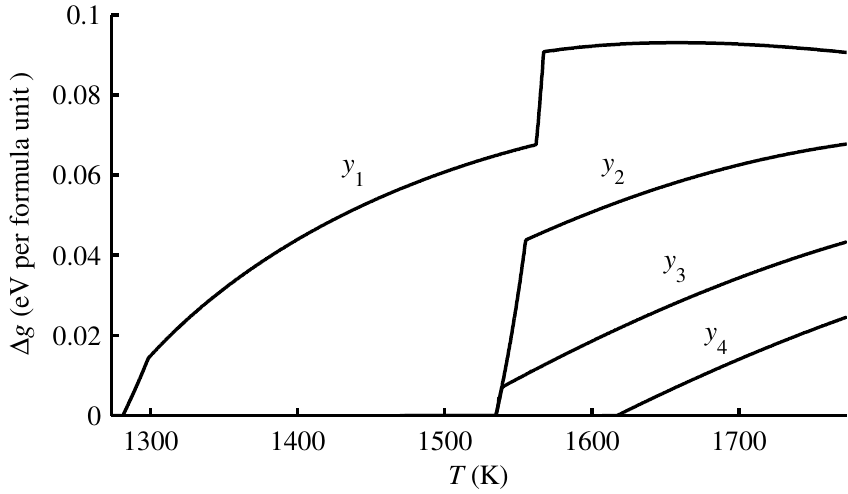}
	\end{center}
	\caption{\label{fig:VC_Deltag}$\Delta g_{\text{(V,W,Co)C}_x}$ as function of temperature $T$ in a W-C-Co-V system in equilibrium with WC and graphite. $y_i$ denotes the V/Co atomic ratio and $y_1 = 6.4 \cdot 10^{-3}$, $y_2 = 2.6 \cdot 10^{-2}$, $y_3 = 5.1 \cdot 10^{-2}$ and $y_4 = 7.7 \cdot 10^{-2}$.}
\end{figure}

In Figure~\ref{fig:VC_Deltag}, the negative driving force of nucleation $\Delta g_{\text{(V,W,Co)C}_x}$ for a $\text{(V,W,Co)C}_x$ phase of optimal composition with respect to Gibbs energy in a W-C-Co-V system in equilibrium with WC and graphite is given as a function of temperature for varying V/Co atomic ratio $y_i$. For low temperatures, $\Delta g_{\text{(V,W,Co)C}_x} = 0$ for all considered doping levels, meaning that stable $\text{(V,W,Co)C}_x$ precipitates exist at these chemical conditions. The sharp increase of $\Delta g_{\text{(V,W,Co)C}_x}$ at temperatures around 1550~K is due to the melting of the binder phase, which significantly increases its solubility of V. At the doping levels $y_2$ and $y_3$ of Figure~\ref{fig:VC_Deltag}, the thermodynamic calculation (which lacks any model for interface stabilization) predicts that all $\text{(V,W,Co)C}_x$ phase would dissolve as the binder melts. A discontinuous change in the slope of $\Delta g_{\text{(V,W,Co)C}_x}$ is also found near $1300 \ \text{K}$ for $y_1$. This change is associated with the transition of the solid Co binder from its ferro- to paramagnetic phase. We conclude that for relevant doping concentrations, $\Delta g_{\text{(V,W,Co)C}_x}$ varies between 0 and 0.1~eV (corresponding to 0 and $0.22 \ \text{J/m}^2$ per layer of $\text{(V,W,Co)C}_x$).

We can now summarize our results for the propensity of thin film formation. In Figure~\ref{fig:VC_stab}, the solid line corresponds to the result from Eq.~(\ref{eq:gamma_film}) for VC. Here, we set $\Delta g_\text{VC} = 0.06 \ \text{eV}$, which is a reasonable choice based on the results from our thermodynamic modeling at relevant sintering temperatures. With this value of $\Delta g_\text{VC}$, Eq.~(\ref{eq:N_limit}) suggests that stable films of thickness less than $N_\text{limit} = 7$ may exist.

In the interface calculations, we have approximated the film carbide phase with a stoichiometric VC phase. The real film may well have a composition similar to the bulk $\text{(V,W,Co)C}_x$ phase, though finding the optimal composition would require numerous large interface calculations. To assess the effect of composition, we have performed a calculation for a mixed $(\text{V}_{0.75},\text{W}_{0.25}) \text{C}$ phase, whose composition is close to the experimentally measured one of bulk $\text{(V,W,Co)C}_x$ \cite{FrMa08}. We obtain $\gamma_{(\text{V}_{0.75},\text{W}_{0.25})\text{C}/\text{Co}} = -0.15 \ \text{J/m}^2 < \gamma_\text{VC/Co}$ \cite{onlinesupp09} and, therefore, we expect that for the true composition, the interfacial effect will further stabilize the thin film.

\subsection{Explicit atomistic layer-by-layer treatment}

The expression for $\gamma_\text{film}$ in Eq.~(\ref{eq:gamma_film}) is only valid if the film is sufficiently thick. For thinner films, the details of the atomic layer-by-layer structure has to be taken into account. We have therefore explicitly modeled the atomic layer-by-layer structure of thin films of cubic VC at the WC/Co interface. The number of possible stacking sequences of VC in the WC/VC/Co interface is large, but can be decreased by omitting obviously high-energetic configurations. In total, we have considered 76 different VC stacking sequences with six corresponding Co stackings for each of these \cite{onlinesupp09}.

\begin{figure}
	\begin{center}
		\includegraphics{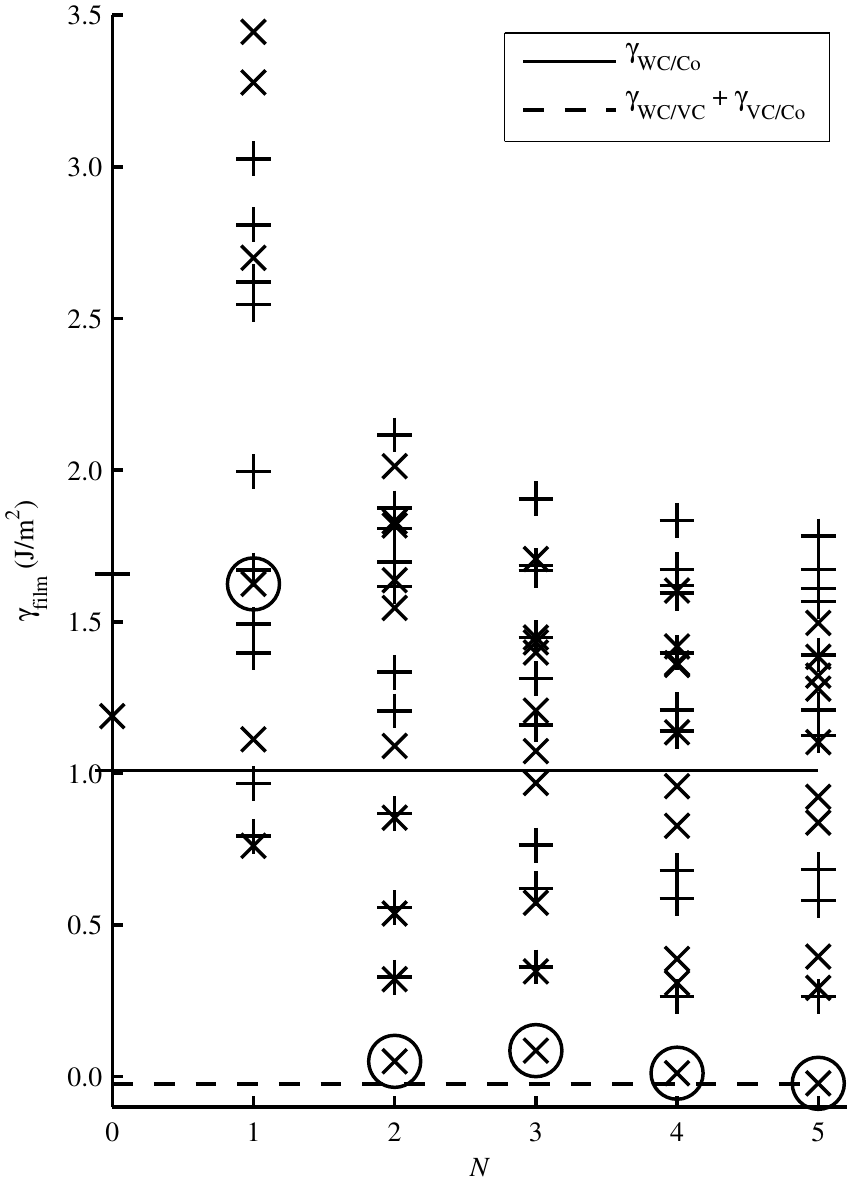}
	\end{center}
	\caption{\label{fig:VC_conv}Interface energy $\gamma_\text{film}$ as a function of the number of V layers $N$ in a cubic VC film in the WC/Co interface for different stackings of VC. + (x) denotes a C-terminated (metal-terminated) carbide/Co interface. The stacking sequence that corresponds to the minimal WC/VC interface energy is circled. No energetic cost per layer to build the thin film is added ($\Delta \bar{g}_\text{VC} + \bar{e}_\text{VC} = 0$).}
\end{figure}

In Figure~\ref{fig:VC_conv}, the energy of a WC/Co interface containing a VC film is given as a function of the number of metallic V layers in the film for different VC stacking sequences. The results for the stackings corresponding to the minimal energy in the WC/VC interface described earlier are circled. In the plot, we show the result without adding any contribution from the energetic cost per layer to build the thin film ($\Delta \bar{g}_\text{VC} + \bar{e}_\text{VC} = 0$). The result should then converge to $\gamma_\text{WC/VC} + \gamma_\text{VC/Co}$ for sufficiently thick films. We find that the result is close to the asymptotic value already for two V layers, while the result for one V layer deviates both energetically and structurally with $\ldots \text{C}_B \text{W}_A \text{C}_B \text{V}_C \text{Co} \ldots$ being the optimal stacking in this case. The optimal structure in the case of two V layers is depicted in Figure~\ref{fig:interface}.

\begin{figure}
	\begin{center}
		\includegraphics{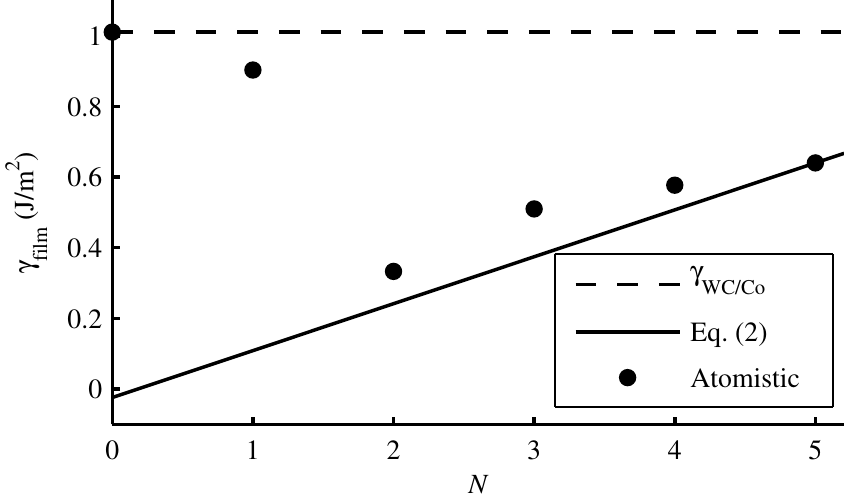}
	\end{center}
	\caption{\label{fig:VC_stab}Interface energy $\gamma_\text{film}$ as a function of the number of V layers $N$ in a cubic VC film in the WC/Co interface. $\Delta g_\text{VC}$ is set to $0.06 \ \text{eV}$ per formula unit, which is an appropriate value for relevant doping conditions and sintering temperatures. A thin film of two V layers will be the most stable configuration for a rather wide interval of $\Delta g_\text{VC}$, $0.01 \ \text{eV} \lesssim \Delta g_\text{VC} \lesssim 0.2 \ \text{eV}$.}
\end{figure}

Our final results are presented as circles in Figure~\ref{fig:VC_stab}, where we again set $\Delta g_\text{VC} = 0.06 \ \text{eV}$. To calculate $\gamma_\text{film}(N)$, we take for each number of V layers $N$ the minimum energy among the stackings and terminations presented in Figure~\ref{fig:VC_conv} and add $N \left(  \Delta \bar{g}_\text{VC} +  \bar{e}_\text{VC} \right)$. The fast convergence of the interface energies toward the asymptotic value of Eq.~(\ref{eq:gamma_film}) implies that a thin film with two V layers will be the most stable configuration for a rather wide interval of $\Delta g_\text{VC}$, $0.01 \ \text{eV} \lesssim \Delta g_\text{VC} \lesssim 0.2 \ \text{eV}$. For no value of $\Delta g_\text{VC}$ will one V layer be the most stable configuration. With the values used in Figure~\ref{fig:VC_stab}, we find that the interface with a VC film is favored by $0.68 \ \text{J/m}^2$ compared to the interface without segregated VC. This value lies well outside our estimated error bound of $\pm 0.35 \ \text{J/m}^2$ indicating that in equilibrium, the WC(0001) surface is indeed covered by VC.

Our prediction of a few stable V layers agrees very well with HREM imaging \cite{LaHaThLa02,LaThHaTh03,LaHaThLo04,LaLoDo04}, where thin films on WC(0001) in VC-doped material are seen to have a thickness of two and sometimes a few additional metallic layers. An interpretation of the additional layers is that they precipitate as $\Delta g_\text{VC} \rightarrow 0$ during cooling. The VC-covered surface is also a likely nucleation site for further growth of the VC phase, which sometimes builds up as a precipitate on WC(0001) during cooling \cite{LaHaThLa02,LaHaThLo04}.

\subsection{Grain growth inhibiting effect}

It is generally believed that the rate limiting step of grain growth in WC-Co systems is the dissolution and/or reprecipitation reactions occuring at WC/Co interfaces. Several authors have therefore attributed the growth inhibiting effect of V to a proposed interference with these reactions either through blocking active growth centers at the WC surface \cite{BoScLu92,KaTeHa04} or through the presence of segregated V at the WC surface \cite{JaYaIkSaTaOkTa98,LaThHaTh03}, which would hinder W diffusion into or out of the WC grain. Our prediction that ultrathin VC films may be present and stable also at liquid-phase sintering temperatures is supportive of the latter explanation.

\section{Conclusions}

We have developed an atomistic model for understanding thin film formation at interfaces in bulk composite materials. We combine extensive density functional theory (DFT) calculations for interfaces and thin films with thermodynamic modeling techniques for multicomponent systems. The latter technique is used to obtain relevant chemical potentials for the various components as function of temperature and doping conditions.

The theory is applied to thin film formation in VC-doped WC-Co cemented carbides. It is predicted that under realistic doping conditions, ultrathin films may exist at high temperature liquid phase sintering conditions where most of the grain growth occurs.

By explicit DFT modeling of the thin films as function of the number of atomic layers, we find that the most stable configuration consists of only two V layers over a large temperature range and we show how the atomistic picture approaches the thick film limit with increasing number of layers.

The films, whose corresponding bulk phase is thermodynamically unstable, are stabilized by interfacial effects. Compared with the original interface in the absence of a thin film, there is an energy gain associated with the creation of the two sub-interfaces involved in the thin film formation, which more than compensates for the energetic cost of the thin film itself.

In conclusion, our results explain the formation of VC films in WC/Co interfaces. The presented method can be applied to theoretically assess the stability of other carbide films in WC/Co interfaces.

\section*{Acknowledgments}

Financial support from the Swedish Research Council, Sandvik and Seco Tools is gratefully acknowledged. Computations have been performed on SNIC resources. We thank Susanne Norgren, Sandvik Tooling, for providing thermodynamic calculations.

\bibliographystyle{tPHL}

\markboth{Sven A. E. Johansson and G{\"o}ran Wahnstr{\"o}m}{}

\pagebreak 




\section*{Supplementary material}

# Theory of ultrathin films at metal-ceramic interfaces - supplementary material

Sven A. E. Johansson and Göran Wahnström
*Department of Applied Physics, Chalmers University of Technology, SE-412 96 Göteborg, Sweden*

## COMPUTATIONAL DETAILS

We employ DFT as implemented in the Vienna *ab-initio* simulation package (VASP) [1]. The exchange-correlation functional is approximated in the Perdew-Burke-Ernzerhof (PBE) scheme [2]. The plane-wave pseudopotential method with projector augmented wave (PAW) [3, 4] potentials is used. The plane-wave energy cutoff is set to 400 eV in all calculations. Partial occupancies are set with the method of Methfessel-Paxton [5] of first order with a smearing parameter of 0.05 eV. Atomic relaxations are performed until all atomic forces are smaller than 0.02 eV/Å.

We consistently use non-spinpolarized calculations, which effectively means that all considered phases are treated as paramagnetic. Although ferromagnetic at low temperatures, Co is paramagnetic above its Curie temperature of 1396 K [6]. The Curie temperature is lowered by solutes, and for Co in equilibrium with WC and graphite, it is around 1324 K (from the model of [6]), which is below the temperatures of concern in the current paper.

## INTERFACE CALCULATIONS

In the interface calculations, the periodic supercell contains two interfacing slabs of the corresponding phases and at least 14 Å of vacuum. This setup allows calculation of energies for individual stacking sequences in the WC (0001) /VC (111) interface, which is not possible in a WC/VC slab setup without vacuum, since the two interfaces formed in such model would be inequivalent due to the *ABC* stacking of the VC phase. The carbide phases are modeled as slabs of $5+4$ atomic layers and the Co phase as a slab of 8 atomic layers. Analogous geometries are used for interfaces containing films of a secondary phase.

For the WC and VC compounds, the smallest possible cell size ($1\times 1$ atoms in the (111) or (0001) planes) is used, where the Brillouin zone is sampled with a $\Gamma$-centered grid with a division of $13 \times 13 \times 1$ points. For the calculation of $\gamma_{(V_{0.75},W_{0.25})C/Co}$, a cell of $2 \times 2$ atoms in the (111) plane is used with a $k$-point sampling of $7 \times 7 \times 1$. Each (111) plane contains 3 V and 1 W atoms, and the placement of W atoms in successive (111) planes is chosen so that W-W interatomic distances are maximized. For bulk calculations, $k$-point sampling is done consistently with the interface calculations. Considering all technical errors, interface energies are hereby converged to a maximal error of 0.015 eV per interfacial unit corresponding to 0.03 J/m$^2$.

The energies described in Tables I, II, III and IV are taken as total energy differences from DFT calculations. As an example,

$$\gamma_{\text{WC/Co}} = \frac{1}{A}\left(E - N_\text{W}\mu_\text{W} - N_\text{C}\mu_\text{C} - N_\text{Co}\mu_\text{Co}\right) - \sigma_\text{Co} - \sigma_\text{WC}, \quad (1)$$

where $E$ is the total energy of the interface system containing $N_\text{W}$ ($N_\text{C}$, $N_\text{Co}$) W (C, Co) atoms. $\mu_\text{W}$, $\mu_\text{C}$ and $\mu_\text{Co}$ are calculated in separate bulk calculations. Due to our choice of geometrical model, which contains vacuum, the surface energies of WC and Co, $\sigma_\text{WC}$ and $\sigma_\text{Co}$, obtained in separate surface calculations are also subtracted. All calculations are done in the same strain state corresponding to fixed lattice parameters in the $(0001)_\text{WC} \parallel (111)_\text{VC} \parallel (111)_\text{Co}$ plane at the calculated equilibrium lattice parameter of WC. Thus, the strain energy in VC and Co is canceled in the calculation of interface energies.

When constructing a WC/Co interface containing a VC film, several stacking sequences are possible for each number of layers of VC. From the values in Table I, we conclude that all stacking sequences that contain two subsequent layers of C or two layers of atoms placed on top of each other (e. g. $\ldots W_A \mid V_A \ldots$) can be disregarded due to their high energy. Furthermore, we only study stacking sequences that are compatible with a nacl stacking. The remaining number of stacking sequences is still rather large. Counting the number of possible stacking sequences for one metallic layer in the film, we find 16 different configurations. For two and more metallic layers, 20 VC stackings are considered. For every VC film stacking, six different stackings of Co on top of the WC+VC slab are modeled in order to give a mean value for the metal-ceramic interface energy. Up to four metallic layers in the film have been explicitly modeled, which gives a total of 456 DFT calculations.

---

TABLE I: Interface energy for different stackings in the WC (0001) /VC (111) interface. Energies are given in J/m$^2$.

| Stacking sequence | $\gamma$ |
|---|---|
| $\ldots W_A C_B W_A \mid V_A C_C V_B C_A \ldots$ | 3.79 |
| $\ldots W_A C_B W_A \mid V_A C_B V_C C_A \ldots$ | 3.87 |
| $\ldots W_A C_B W_A \mid V_B C_C V_A C_A \ldots$ | 1.10 |
| $\ldots W_A C_B W_A \mid V_B C_A V_C C_B \ldots$ | 1.27 |
| $\ldots W_A C_B W_A \mid V_C C_B V_A C_C \ldots$ | 1.32 |
| $\ldots W_A C_B W_A \mid V_C C_A V_B C_C \ldots$ | 1.49 |
| $\ldots W_A C_B W_A \mid C_A V_C C_B V_A \ldots$ | 5.20 |
| $\ldots W_A C_B W_A \mid C_A V_B C_C V_A \ldots$ | 5.19 |
| $\ldots W_A C_B W_A \mid C_C V_B C_A V_C \ldots$ | 1.38 |
| $\ldots W_A C_B W_A \mid C_C V_A C_B V_C \ldots$ | 0.92 |
| $\ldots C_B W_A C_B \mid V_A C_C V_B C_A \ldots$ | -0.03 |
| $\ldots C_B W_A C_B \mid V_A C_B V_C C_A \ldots$ | 0.39 |
| $\ldots C_B W_A C_B \mid V_B C_C V_A C_B \ldots$ | 3.04 |
| $\ldots C_B W_A C_B \mid V_B C_A V_C C_B \ldots$ | 3.03 |
| $\ldots C_B W_A C_B \mid V_C C_B V_A C_C \ldots$ | 0.83 |
| $\ldots C_B W_A C_B \mid V_C C_A V_B C_C \ldots$ | 0.29 |
| $\ldots C_B W_A C_B \mid C_A V_C C_B V_A \ldots$ | 7.81 |
| $\ldots C_B W_A C_B \mid C_A V_B C_C V_A \ldots$ | 6.28 |
| $\ldots C_B W_A C_B \mid C_B V_C C_A V_B \ldots$ | 4.13 |
| $\ldots C_B W_A C_B \mid C_B V_A C_C V_B \ldots$ | 4.09 |
| $\ldots C_B W_A C_B \mid C_C V_B C_A V_C \ldots$ | 8.22 |
| $\ldots C_B W_A C_B \mid C_C V_A C_B V_C \ldots$ | 9.72 |

TABLE II: Interface energy for different stackings in the WC (0001) /Co (111) interface. Energies are given in J/m$^2$.

| Stacking sequence | $\gamma$ |
|---|---|
| $\ldots W_A C_B W_A \mid Co_A Co_B Co_C Co_A \ldots$ | 2.59 |
| $\ldots W_A C_B W_A \mid Co_A Co_C Co_B Co_A \ldots$ | 2.53 |
| $\ldots W_A C_B W_A \mid Co_B Co_A Co_C Co_B \ldots$ | 0.07 |
| $\ldots W_A C_B W_A \mid Co_B Co_C Co_A Co_B \ldots$ | 0.35 |
| $\ldots W_A C_B W_A \mid Co_C Co_A Co_B Co_C \ldots$ | 0.64 |
| $\ldots W_A C_B W_A \mid Co_C Co_B Co_A Co_C \ldots$ | 0.94 |
| Mean | 1.19 |
| $\ldots C_B W_A C_B \mid Co_A Co_B Co_C Co_A \ldots$ | -0.02 |
| $\ldots C_B W_A C_B \mid Co_A Co_C Co_B Co_A \ldots$ | 0.22 |
| $\ldots C_B W_A C_B \mid Co_B Co_A Co_C Co_B \ldots$ | 3.27 |
| $\ldots C_B W_A C_B \mid Co_B Co_C Co_A Co_B \ldots$ | 3.16 |
| $\ldots C_B W_A C_B \mid Co_C Co_A Co_B Co_C \ldots$ | -0.21 |
| $\ldots C_B W_A C_B \mid Co_C Co_B Co_A Co_C \ldots$ | -0.36 |
| Mean | 1.01 |

TABLE III: Interface energy for different stackings in the VC (111) /Co (111) interface. Energies are given in J/m$^2$.

| Stacking sequence | $\gamma$ |
|---|---|
| $\ldots C_A V_B C_C V_A \mid Co_A Co_B Co_C Co_A \ldots$ | 0.62 |
| $\ldots C_A V_B C_C V_A \mid Co_A Co_C Co_B Co_A \ldots$ | 0.68 |
| $\ldots C_A V_B C_C V_A \mid Co_B Co_A Co_C Co_B \ldots$ | -0.51 |
| $\ldots C_A V_B C_C V_A \mid Co_B Co_C Co_A Co_B \ldots$ | -0.24 |
| $\ldots C_A V_B C_C V_A \mid Co_C Co_A Co_B Co_C \ldots$ | -0.36 |
| $\ldots C_A V_B C_C V_A \mid Co_C Co_B Co_A Co_C \ldots$ | -0.16 |
| Mean | 0.00 |
| $\ldots V_B C_C V_A C_B \mid Co_A Co_B Co_C Co_A \ldots$ | -0.99 |
| $\ldots V_B C_C V_A C_B \mid Co_A Co_C Co_B Co_A \ldots$ | -0.73 |
| $\ldots V_B C_C V_A C_B \mid Co_B Co_A Co_C Co_B \ldots$ | 3.45 |
| $\ldots V_B C_C V_A C_B \mid Co_B Co_C Co_A Co_B \ldots$ | 3.27 |
| $\ldots V_B C_C V_A C_B \mid Co_C Co_A Co_B Co_C \ldots$ | -1.53 |
| $\ldots V_B C_C V_A C_B \mid Co_C Co_B Co_A Co_C \ldots$ | -1.71 |
| Mean | 0.29 |

TABLE IV: Interface energy for different stackings in the $(V_{0.75}, W_{0.25})C(111)/Co(111)$ interface. Energies are given in J/m$^2$.

| Stacking sequence | $\gamma$ |
|---|---|
| $\ldots C_A (V,W)_B C_C (V,W)_A \mid Co_A Co_B Co_C Co_A \ldots$ | 0.64 |
| $\ldots C_A (V,W)_B C_C (V,W)_A \mid Co_A Co_C Co_B Co_A \ldots$ | 0.71 |
| $\ldots C_A (V,W)_B C_C (V,W)_A \mid Co_B Co_A Co_C Co_B \ldots$ | -0.72 |
| $\ldots C_A (V,W)_B C_C (V,W)_A \mid Co_B Co_C Co_A Co_B \ldots$ | -0.45 |
| $\ldots C_A (V,W)_B C_C (V,W)_A \mid Co_C Co_A Co_B Co_C \ldots$ | -0.63 |
| $\ldots C_A (V,W)_B C_C (V,W)_A \mid Co_C Co_B Co_A Co_C \ldots$ | -0.42 |
| Mean | -0.15 |
| $\ldots (V,W)_B C_C (V,W)_A C_B \mid Co_A Co_B Co_C Co_A \ldots$ | -1.03 |
| $\ldots (V,W)_B C_C (V,W)_A C_B \mid Co_A Co_C Co_B Co_A \ldots$ | -0.78 |
| $\ldots (V,W)_B C_C (V,W)_A C_B \mid Co_B Co_A Co_C Co_B \ldots$ | 3.25 |
| $\ldots (V,W)_B C_C (V,W)_A C_B \mid Co_B Co_C Co_A Co_B \ldots$ | 3.09 |
| $\ldots (V,W)_B C_C (V,W)_A C_B \mid Co_C Co_A Co_B Co_C \ldots$ | -1.55 |
| $\ldots (V,W)_B C_C (V,W)_A C_B \mid Co_C Co_B Co_A Co_C \ldots$ | -1.72 |
| Mean | 0.21 |



\end{document}